\input{aipcheck}

\documentclass[
    ,final            
  ]
  {aipproc}

\layoutstyle{6x9}

\begin{document}

\title{Plerionic Supernova Remnants}

\author{Samar Safi-Harb}{
  address={Department of Physics and Astronomy, University of Manitoba, Winnipeg, MB R3T 2N2, Canada}
   ,altaddress={Canada Research Chair; samar@physics.umanitoba.ca} 
}

\begin{abstract}

Plerions represent ideal laboratories for the search for neutron stars, the study of their relativistic winds, and their interaction with their surrounding supernova ejecta and/or the interstellar medium. As well, they are widely believed to represent efficient engines for particle acceleration up to the knee of the cosmic ray spectrum (at about 10$^{15}$ eV). Multi-wavelength observations from the radio to the highest TeV energies, combined with modelling, have opened a new window to study these objects, and particularly shed light on their intrinsic properties, diversity, and evolution. High-resolution X-ray observations are further revealing the structure and sites for shock acceleration. The missing shells in the majority of these objects remain puzzling, and the presence of plerions around highly magnetized neutron stars is still questionable. I review the current status and statistics of observations of plerionic supernova remnants (SNRs), highlighting combined radio and X-ray observations of a growing class of atypical, non Crab-like, plerionic SNRs in our Galaxy. I will also briefly describe the latest developments to our high-energy SNRs catalogue recently released to the community, and finally highlight the key questions to be addressed in this field with future high-energy missions, including Astro-H in the very near future.

\end{abstract}

\maketitle

\section{Plerionic Supernova Remnants}

Supernova Remnants (SNRs) are generally classified under three categories: Shells, Filled-centre, or Composites\footnote{see Dave Green's catalogue, http://www.mrao.cam.ac.uk/surveys/snrs/,
for a compilation and classification of the Galactic Supernova Remnants based on radio studies.}.
Filled-centre SNRs have been referred to in the literature  as `plerions' or  Pulsar Wind Nebulae (PWNe), and Composite-type SNRs are classified as either Plerionic Composites (those powered by plerions) or thermal Composites (also referred to as `Mixed-Morphology' SNRs).
The term `plerion' originates from the ancient greek word pleres ($\pi \lambda \eta \rho \eta s$) \cite{weilerpanagia78}
which means `filled'; the centrally filled morphology arising from the non-thermal emission powered by a neutron star.

Typified by the Crab nebula, plerions are bubbles of relativistic particles inflated by the pulsar's relativistic wind.
As such they are  `calorimeters' for the study of neutron stars and, in the absence of a pulsar detection, excellent pathfinders for pulsar discovery.
They allow us to probe the interaction between the neutron stars' relativistic winds and their surroundings, 
which consist of the supernova ejecta at the earlier stages of SNR evolution, the shocked interstellar matter or supernova blast wave as the neutron star encounters the expanding SNR shell,
and the interstellar medium (ISM) after the neutron star has left its hosting SNR.
Plerions seed the Galaxy with highly energetic particles and strong magnetic fields, and are thus considered efficient engines for cosmic ray acceleration up to TeV energies.

 Plerions are observed from radio to the highest energy gamma-rays, and  are {\it generally} characterized by the following properties:
1) an increase in the brightness towards the centre, unlike shells which have a limb-brightened morphology;
2) their X-ray size is generally shorter than
their radio and optical size, due to the smaller synchrotron lifetimes of the
higher energy photons; 
3) a flat radio spectral index, $\alpha$ = 0--0.3 ($S_{\nu}$$\sim$$\nu^{\-\alpha}$) with
 a steeper index in the X-ray due to synchrotron and radiation losses and with an average photon index
$\Gamma$~=~$\alpha$+1$\sim$2; and
4) they are highly polarized with a preferentially radial magnetic
field whose strength varies from a few $\mu$G -- mG.

Modern X-ray telescopes, combined with numerical simulations, have greatly advanced our understanding of plerions \cite{gaenslerslane2006, KP2008, KP2010, bucciantini11}. In particular, they have allowed us to 
1) zoom in on the interaction of the relativistic pulsar wind with its surroundings, showing evidence of torus and jet-like structures successfully reproduced in simulations,
and 2) map the photon index across the plerion, answering questions related to the energetics of the powering engine and the wind magnetization parameter at the termination shock.
Correlating the X-ray with radio and gamma-ray studies sheds light on the magnetic field configuration and strength, the conditions of the confining ambient medium, the age and the distance to the object.
While their radio to X-ray emission arises from synchrotron radiation from relativistic particles injected by the neutron star.
 their gamma-ray emission arises, in the leptonic model, from Inverse Compton scattering (with the seed photons coming from the synchrotron emitting relativistic particles in the pulsar wind, or from the cosmic microwave background, dust and starlight), and in the hadronic model from pion decay.
 
\subsection{Evolution of Plerions}
The evolution of a plerion in an SNR has been studied by several authors, including \cite{blondin2001, bucciantini2003, vanderswaluw2004}.
Initially plerions expand in the freely moving ejecta of the SNR. As the SNR evolves from the free expansion phase into the Sedov-Taylor phase, a reverse shock develops that runs back into the expanding ejecta and can eventually reach the central plerion
on a timescale of $\sim$10 kyr. This is followed by a crushing and re-expansion phases of the plerion.
This action displaces the plerion from the pulsar, leaving a `relic' nebula which can still be detected in the radio.
The `crushing' phase is subject to Rayleigh-Taylor instabilities that result in the mixing of thermal (ejecta) and non-thermal (PWN) fluids. 
Furthermore, asymmetries in the position of the plerion relative to the pulsar and explosion site arise from asymmetries in the surrounding ISM.
Meanwhile the continuous injection of relativistic particles from the neutron star forms an X-ray nebula surrounding it and that is offset from the relic nebula.
The pulsar's motion inside the SNR further causes this wind bubble of freshly injected particles to move with it through the SNR interior.
When its motion becomes supersonic, it forms a bow-shock nebula.

The best and most studied example for a remnant displaying both a bow-shock and relic nebulae interacting with an asymmetrical reverse shock  is the Vela-X region in the $\sim$11-kyr old SNR Vela.
The complex morphology and properties of the nebula in the radio \cite{dodson2003}, X-rays \cite{MO95, pavlov2003, lamassa08} and gamma-rays \cite{aharonian2006} suggest
that a reverse shock has displaced the original plerion (Vela X) from the pulsar.
While Vela X is believed to be a relic nebula, the Vela pulsar has created a new nebula in its vicinity that clearly has a bow-shock morphology \cite{blondin2001, CR2011}.
Another example of an offset plerion likely interacting with the reverse shock is in the Composite-type SNR G327.1$-$1.1 \cite{temim2009},
 estimated to have an age of $\sim$18~kyr and recently discovered in TeV gamma-rays with H.E.S.S. \cite{acero12}.
The overall morphology is also similar to that seen in other offset TeV PWNe, such as HESS J1825$-$139 \cite{aharonian2006b}.

 \section{Statistics and the first X-ray plus gamma-ray SNR catalogue}

As of October 2012, we know of 309 Galactic SNRs; see \cite{FSSH2012} for a 
an up-to-date (and first of its kind) catalogue\footnote{http://www.physics.umanitoba.ca/snr/SNRcat} of X-ray and gamma-ray observations 
of all Galactic SNRs. Our high-energy catalogue complements Dave Green's catalogue based on radio studies.
 Out of these 309 remnants, $\sim$90 SNRs contain plerions or plerionic candidates,  about
30 of which lack shells (including the Crab). These will be referred to as `naked' PWNe.
 Out of the 70 Galactic TeV known sources, about half are identified as PWNe or PWN candidates.
 In the majority of the latter class, the TeV emission is found to be offset from the X-ray peak suggestive of `relic' PWNe (like for Vela-X).
 The TeV detections are particularly useful for studying low-magnetic field plerions, since these could be X-ray faint but gamma-ray bright.

\section{Growing Class of `unusual' Plerions}

For over three decades, our view of plerions has been shaped by the study
of the young and bright Crab nebula. However, there are several plerions whose properties
differ from the Crab, hinting to a diversity of PWNe which include many previously missed evolved objects. 
In the following, I highlight
some interesting cases that we have been studying using coordinated radio and X-ray observations,
and that need to be further investigated in gamma-rays and with deeper X-ray observations.

\begin{figure}
 \includegraphics[height=.5\textheight]{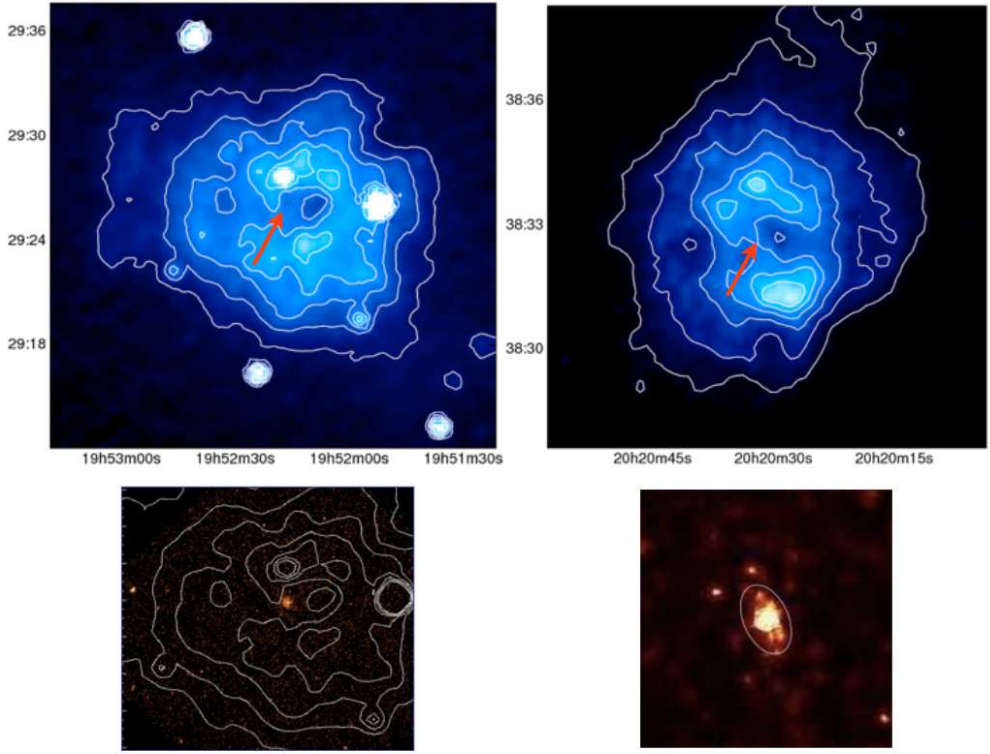}
 \caption{Top: The radio images of the plerionic SNRs DA~495 (left) and G76.9+1.1 (right), with the arrow pointing to the location of the X-ray source discovered with $Chandra$.
 The bottom panels show the $Chandra$ images of the central regions of the SNRs highlighting the discovery of compact X-ray nebulae, inside the depression seen in the radio, surrounding pulsar candidates.
 For  DA~495, the X-ray nebula is $\sim$40$^{\prime\prime}$ in diameter surrounding the point-like source CXOU J202221.68+384214. 
 For G76.9+1.0, the X-ray image has a scale size of 1$^{\prime}$$\times$1$^{\prime}$ and the central ellipse shows the compact nebula with a size of 8$^{\prime\prime}$.3$\times$5$^{\prime\prime}$.1
 surrounding PSR~J2022+3842.
Adapted from figures and data published in \cite{Arzoumanian2008}, \cite{Kothes2008}, and \cite{Arzoumanian2011}.}
 \end{figure}

\subsection{DA~495 (G65.7+1.2) and G76.9+1.0}

The SNRs DA~495 (G65.7+1.2) and G76.9+1.0 (see Fig.~1) are unusual due to the following reasons: 
1) unlike shell-type SNRs, they are not limb-brightened, but their non-thermal flux falls off away from centre, implying a neutron star as a central engine;
2) their radio spectrum has a very steep spectral index, $\alpha$ $\sim$ 0.6, which is more typical of Shell-type remnants; 
3) they are large and have a low-frequency spectral break in the radio; and
4) in polarized emission, they have a strong axisymmetry and a double-lobed morphology.

\begin{table}[tbh]
\begin{tabular}{lrr|rr}
\hline
 Component & DA495 Nebula & pulsar candidate & G76.9+1.0 Nebula & PSR~J2022+3842 \\
\hline
$N_{\rm H}$ & $\leq$0.4$\times$10$^{22}$~cm$^{-2}$ & ... & 1.6$\pm$0.3$\times$10$^{22}$~cm$^{-2}$ & ... \\
$\Gamma$/kT (MK) & 1.6 & 2.5 & 1.4 & 1.0$\pm$0.2\\
$L_x$\tablenote{in units of erg~s$^{-1}$, and at a distance of 1~kpc for DA~495 and 10 kpc for G76.9+1.0} & 3.3$\times$10$^{31}$ & $\sim$10$^{31}$ & 5.6$\times$10$^{32}$  & 7$\times$10$^{33}$ \\
\hline
\end{tabular}
\caption{Summary of the properties of the plerion and pulsar candidate/pulsar in the SNRs DA~495 and G76.9+1.0 \cite{Arzoumanian2008, Arzoumanian2011}. All models are power-law except for the point source in DA~495 which is fitted with a blackbody model.}
\end{table}

While their nature as SNRs has been questioned in the past, $Chandra$ observations have allowed the discovery of compact non-thermal nebulae surrounding point-like sources, sitting in the heart of the radio depression. The morphology and spectral properties of both the nebulae and point-like sources (see Table~1) left no doubt about their nature as plerions. Combined X-ray and radio studies have further suggested that these are evolved PWNe with a very low X-ray to radio luminosity and size ratios, suggestive of a high nebular magnetic field~\cite{Kothes2008}.
Follow-up radio pulsation searches failed to detect a pulsar in DA~495, but led to the discovery of a 24 ms pulsar (PSR J2022+3842)  in G76.9+1.0, a result that was confirmed in X-rays with the Rossi X-ray Timing Explorer \cite{Arzoumanian2011}.
While DA~495 hasn't been yet detected in gamma-rays, there has been recently a reported Fermi detection in G76.9+1.0 \cite{nolan12}.

\subsection{CTB~87 (G74.9 +1.2) and G63.7+1.1}

\begin{figure}[tbh]
 \includegraphics[height=.35\textheight]{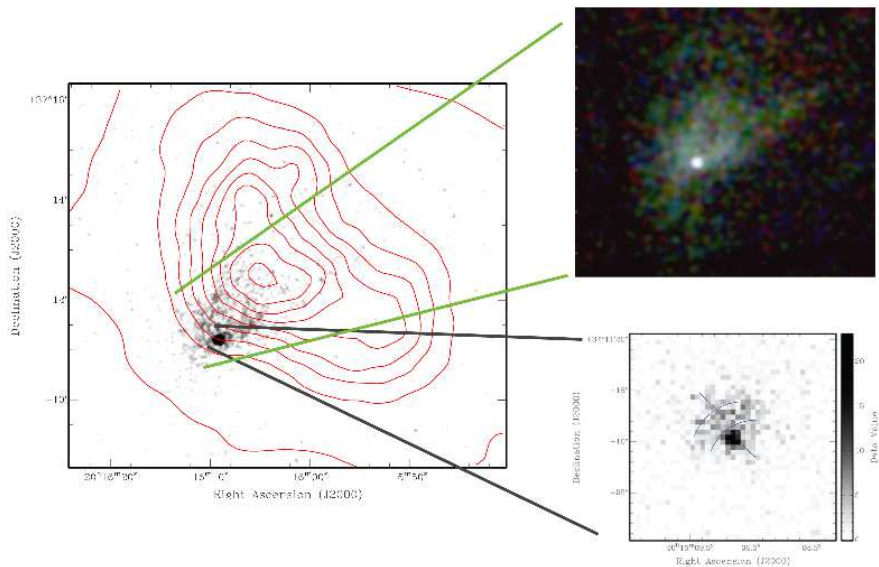}
\caption{Left Panel: The \textit{Chandra} greyscale X-ray image of CTB~87 with the Canadian Galactic Plane Survey radio contours overlaid (in red). The top right panel shows the RGB energy image of the extended diffuse nebula
with a cometary morphology powered by a pulsar candidate. The bottom panel shows a more compact nebula surrounding the putative pulsar with evidence of faint jet-like and torus-like structures. See \cite{MSHK2012} for details.}
\end{figure}

CTB~87 and G63.7+1.1 are two plerionic candidates, previously unexplored in X-rays, but that also have unusual characteristics in radio wavelengths although with a morphology different from the plerions discussed in the previous section.
They both have a centrally filled morphology (as for plerions) but  they are also unusual in that they have 1) a steep radio spectral index ($\alpha$~=~0.4--0.5), 2) a low-frequency spectral break in the radio, and 3) they are unusually large implying evolved PWNe.
Our $Chandra$ and \textit{XMM-Newton} studies unveiled faint X-ray PWNe in both objects, surrounding pulsar candidates, and that are clearly offset from the peak of the radio emission (papers in preparation). In CTB~87 shown in Fig.~2, we find evidence of both a compact nebula with a torus/jet-like structure and a diffuse nebula with a cometary-like morphology \cite{MSHK2012}. The radio and X-ray study suggests that CTB~87 represents an evolved ($\sim$10--20~kyr old) plerion where the radio emission peak corresponds to a relic PWN, similar to what's observed in G327.1--1.1.
Recent gamma-ray observations with VERITAS led to the detection of this source \cite{aliu11}.

\subsection{Plerions around highly magnetized neutron stars?}

\begin{figure}[tbh]
 \includegraphics[height=.3\textheight]{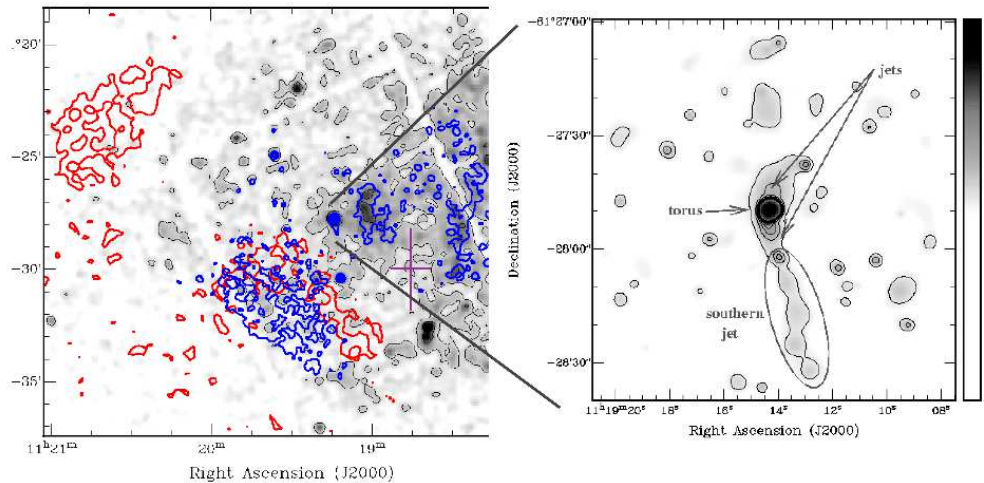}
  \caption{ \emph{Right}: The compact PWN and jet discovered with \textit{Chandra} surrounding the high-magnetic field pulsar J1119$-$6127~\cite{sshkumar08} in the SNR G292.2$-$0.5 shown in the
  left panel. \emph{Left:}   \textit{XMM-Newton} image and contours in greyscale, with the \textit{Chandra} contours overlaid in blue (darker contours). The red contours show the optical/DSS emission
  illustrating the presence of dark clouds in the \textit{eastern} field of G292.2$-$0.5. The purple cross in the \textit{western} field refers to the H.E.S.S. detection in the SNR interior \cite{djannati09}. Figure adapted from figures published in \cite{kumarsshgonz12} and \cite{sshkumar08} and best seen in colour in the electronic version.}
\end{figure}

The question on whether highly magnetized neutron stars power plerions remains to be answered. So far among the known 23 magnetars plus 7 X-ray detected high-magnetic radio field pulsars, 
there are only 4 objects confirmed to power PWNe discovered or detected in X-rays: 
the high-magnetic field pulsars PSR J1119$-$6127 in the SNR G292.2$-$0.5~\cite{sshkumar08, gonzssh03} (see Fig.~3) and
PSR J1846$-$0258 in the SNR Kes~75~\cite{kumarssh08, gavriil08, ng08},
PSR J1819-1458 (classified as a RRAT or Rotating Radio Transient) \cite{rea09}, and the magnetar Swift J1834.9$-$0846~\cite{younes12}.
The PWN claimed to be associated with the anomalous X-ray pulsar 1E 1547.0$-$5408 \cite{bambavink09} was recently proposed to be a dust scattering halo \cite{olausen11}.
Among these, Kes 75 and G292.2$-$0.5 have been detected in gamma-rays with H.E.S.S.\footnote{http://www.mpi-hd.mpg.de/hfm/HESS/pages/home/som/2008/10/}$^,$\footnote{http://www.mpi-hd.mpg.de/hfm/HESS/pages/home/som/2009/11/}.
In G292.2$-$0.5, the H.E.S.S. source was located in the western field of the SNR, between the pulsar and the SNR shell where hard non-thermal X-ray emission was also detected (Fig.~3).
While the origin of the gamma-ray emission is not yet understood, it could be attributed to an offset plerion, where the nebula has been displaced after being crushed by an asymmetric reverse shock
\cite{kumarsshgonz12, djannati09}.

\begin{figure}
 \includegraphics[height=.3\textheight,width=1.0\textwidth]{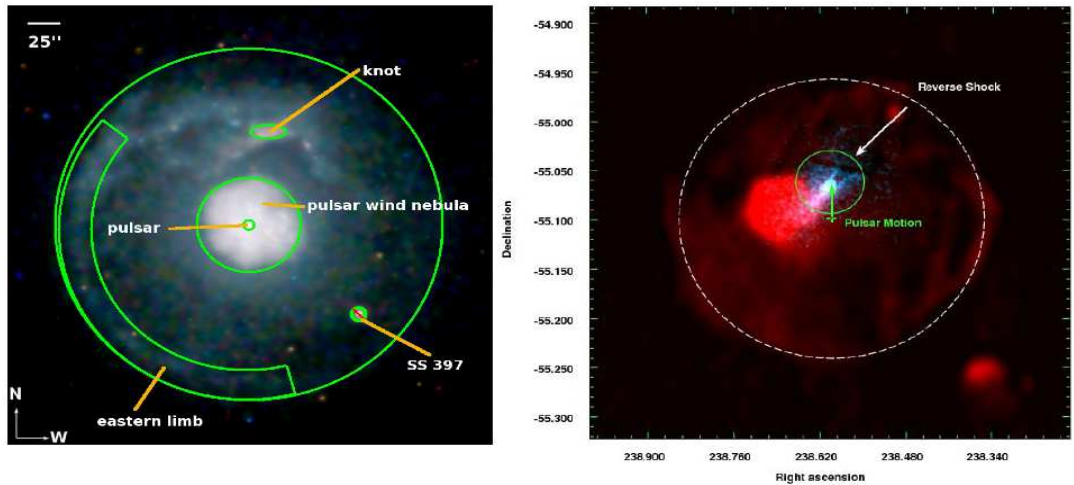}
  \caption{\emph{Left}: The young plerionic SNR G21.5$-$0.9 with $Chandra$ showing evidence for thermal emission from `knots' outside the main PWN component, and a limb-brightened shell to the east. Image credit: \cite{MSSH2010}. \emph{Right}: The composite $Chandra$ (in blue) and radio (in red) image of the Composite-type SNR G327.1--1.1 harbouring an offset plerion recently detected in gamma-rays. Image credit: \cite{temim2009}.}
\end{figure}

\section{Conclusions and future prospects}

The synergy between radio and high-energy observations has been crucial to understanding the emission mechanism in plerions and in addressing
their apparent diversity. In particular, radio plus X-ray observations are allowing the discovery of a new class of plerions which seem to have unusual properties,
but that likely represent evolved objects.
Some of the outstanding questions in this field include:
\begin{itemize} 
\item{Where is the shell around the dozens of `naked' plerions? Is the absence of a shell caused by the nature of the progenitor star creating a wind-blown bubble in which the plerion is expanding? 
Are they born in lower energy supernova explosions?}
\item{Do all high magnetic field pulsars and magnetars power nebulae?  What powers their X-ray emission: rotation, magnetism, both, or some other mechanism?}
\end{itemize}

Deep observations with current observatories as well as targeted multi-wavelength 
observations will address these, and other, questions related to plerion physics and evolution.
In X-rays, we look forward to the launch of the Astro-H mission \cite{takahashi2010} which will have an unprecedented spectral resolution and sensitivity, as well as a broadband coverage in the 0.5--600 keV energy range.
This will allow the search for the long sought SNR shells and thermal plasma in the `naked' plerions, and for the thermal emission expected in evolved plerions crushed by the reverse shock (see Fig.~4).
The unique broadband coverage will also measure their broadband X-ray spectra and, in conjunction with radio and gamma-ray studies, help constrain their magnetic field and emission mechanism (leptonic versus hadronic).
Finally, the synergy between X-ray and gamma-ray observations with current and the new generation observatories (including Fermi,  H.E.S.S. II,  VERITAS, and the Cerenkov Telescope Array in the future)
will allow us to complete the census of Plerionic SNRs in our Galaxy and beyond.

\begin{theacknowledgments}
 The author acknowledges the support of the Canada Research Chairs program, the Natural Sciences and Engineering Research Council of Canada,
the Canadian Institute for Theoretical Astrophysics, the Canadian Space Agency, and the Canada Foundation for Innovation.
She is grateful to the organizers for the invitation and the opportunity to attend a stimulating and well organized conference
in a nice city. She also acknowledges the many contributions to the results presented in this review by her
collaborators particularly Zaven Arzoumanian, Roland Kothes and Tom Landecker; group members
including Heather Matheson, Gilles Ferrand, and Harsha Kumar; as well as the entire active and productive SNR/PWN community.
\end{theacknowledgments}

\bibliographystyle{aipprocl} 

\end{document}